
\magnification=1200
\line{\hskip 4 in UTHEP-236 \hfill}
\line{\hskip 4 in May, 1992 \hfill}
\bigskip
\bigskip
\centerline{\bf REGULARIZATIONS, ANOMALIES AND FERMION NUMBER}
\centerline{\bf NON-CONSERVATION IN CHIRAL GAUGE THEORIES }
\bigskip
\bigskip
\centerline{ Sinya Aoki$^{\circ }$}
\bigskip
\bigskip
\centerline{ Institute of Physics, University of Tsukuba,
Tsukuba, Ibaraki 305, Japan}
\smallskip
\advance
\baselineskip by 0.5
\baselineskip
\bigskip
\bigskip
\centerline{ ABSTRACT}
\bigskip
We study how fermion number conservation fails
in fermion number preserving regularization schemes.
We show that the fermion number have to be carried by
the gauge field configurations with non-zero winding number in this
scheme and this fermion number is not conserved in the presence of
instantons.
We also consider other types of regularization scheme
which have different global symmetries. In particular,
we point out that the fermion number is conserved
in the lattice chiral gauge theories with the Wilson-Yukawa coupling.
\vskip 1.5 in
\vfill
\leftline{$\circ$ email: aoki@ph.tsukuba.ac.jp}
\eject
\def\pl#1#2#3{{\it Phys. Lett.} {\bf B#1}(#2)#3}
\def\np#1#2#3{{\it Nucl. Phys.} {\bf B#1}(#2)#3}
\def\npp#1#2#3{{\it Nucl. Phys.} {\bf B}(Proc. Suppl.){\bf #1}(#2)#3}
\def\pr#1#2#3{{\it Phys. Rev.} {\bf D#1}(#2)#3}
\def\prl#1#2#3{{\it Phys. Rev. Lett.} {\bf #1}(#2)#3}
\def\bar#1{\overline#1}
\def\vac#1{\langle #1 \rangle}
\centerline{I. \ Introduction}
\bigskip

It is well-known[1]
that the sum of the baryon number ($B$) and the lepton
number ($L$) is not conserved in the standard model due to an anomaly
and SU(2) instantons. Recently it is pointed out[2],
however, that the lattice formulation of chiral gauge theories
with the so-called Wilson-Yukawa coupling[3-6]
can not produce this anomaly,
hence $B$ and $L$ are conserved. This fact is rather
surprising since this means that
the structure of the  anomaly could depend on the
way how one regularizes the theory.
In this paper we investigate the relation among the regularizations,
anomalies and the fermion number non-conservation.

Before considering the relation, we first mention subtleties of
regularizations for chiral gauge theories.
We consider a chiral gauge theory without
Higgs fields, whose partition function is given by
$$
Z =\int {\cal D}\psi{\cal D}\bar\psi{\cal D}W_\mu
\exp [ S_0(\psi,\bar\psi, W_\mu)+ S_{reg}(\psi,\bar\psi,W_\mu)  ]
\eqno(1.1)$$
where $\psi$ are fermion fields, $W_\mu$ are gauge fields,
$S_0$ is the classical part of the action, and $S_{reg}$ is
the regulator part of the action.
The problem in the regularization of chiral gauge theories is that
$S_{reg}$ is {\it not} invariant under the gauge transformation
$$ \left\{ \eqalign{
\psi^h  & = ( h_L P_L + h_R P_R ) \psi \cr
\bar\psi^h &  =\bar\psi ( h_L^\dagger P_R + h_R^\dagger P_L ) \cr
W_\mu^h  & = h W_\mu h^\dagger-i (\partial_\mu h)h^\dagger  \cr },
\right. \eqno(1.2)
$$
where $h$ is a gauge transformation function which satisfies
$h^\dagger h = 1$, $ h_X = D^X (h)$ for $X= L$ or $R$, and $D^X$ is some
unitary representation of the gauge group for the fermions.

Using the identity
$$
\int {\cal D}g \exp [ S_{GF} ( \psi^{g^\dagger}, \bar\psi^{g^\dagger},
W_\mu^{g^\dagger} ) ] = 1 \eqno(1.3)
$$
where
$S_{GF}$ is a gauge-fixing functional including the Faddeev-Popov
determinant\footnote{$^1$}{We can take $S_{GF} = 0$ for the lattice
regularization.}, we obtain
$$
Z =\int {\cal D}g {\cal D}\psi {\cal D}\bar\psi {\cal D}W_\mu
\exp [ S_0(\psi, \bar\psi, W_\mu)+ S_{reg}(\psi^g,\bar\psi^g,W_\mu^g)
+ S_{GF} (\psi, \bar\psi, W_\mu )]  . \eqno(1.4)
$$
Since $S_{reg}$ is not gauge invariant, the group-valued field $g$
appears in $S_{reg}$, so that the gauge volume
$\displaystyle \int {\cal D}g$ can not
be factored out. There are two approaches for treating the
scalar field $g$. In the first approach we try to decouple $g$ field
from the renormalized theory by adding local gauge non-invariant
counter terms. This procedure gives a renormalizable and gauge ( or BRST)
invariant renormalized perturbation theory for anomaly free theories.
However, the gauge invariance (1.2) is lost at the regularized level.
We call this approach as the gauge non-invariant scheme.
In the other approach $g$ field is considered to be the Nambu-Goldstone part of
the Higgs field, so that the regularized theory describes a chiral gauge
theory {\it with} the Higgs field. Although the gauge symmetry (1.2) is
lost,
the regularized action is invariant under another local transformation:
$$\left\{ \eqalign {
\psi^h & = ( h_L P_L + h_R P_R ) \psi \cr
\bar\psi^h & =\bar\psi ( h_L^\dagger P_R + h_R^\dagger P_L ) \cr
W_\mu^h & = h W_\mu h^\dagger-i (\partial_\mu h)h^\dagger  \cr
g^h & = g h^\dagger \cr} \right. \eqno(1.5)
$$
which is identical to the gauge transformation of the
gauge-Higgs-fermion system.
However, the theory is not manifestly renormalizable due to the
non-linearity of $g$. We call this approach as the gauge invariant scheme.
The lattice formulation of chiral gauge theories with the
Wilson-Yukawa coupling belongs to this scheme.

If we consider theories without Yukawa couplings,
the calculation of the fermion determinant for background gauge
fields is identical in both schemes
if we take the unitary gauge such that $g = 1$.
In order to interpret the results
we impose the gauge invariance (1.2) for the renormalized theory in the gauge
non-invariant regularization scheme, while the gauge symmetry (1.5),
not (1.2), is relevant in the gauge invariant regularization scheme and
(1.5) is automatically satisfied in this scheme.
\vskip 1cm
\centerline{II. \ Anomalies of Noether Currents }
\vskip 0.5cm

In this section we give a formula by which we can easily obtain
the divergence of the vector and axial-vector Noether currents as well as
their variation under (1.2).

We consider the theory defined by the action:
$$
S =\int dx \sum_{\mu =1}^4 \bar\psi (x)
\gamma^\mu [\partial_\mu + i W_\mu^L(x) P_L +
iW_\mu^R(x) P_R  ] \psi (x) \eqno(2.1)
$$
where $\psi$, $\bar\psi$ are fermion fields,
$P_{L,R}=\displaystyle { 1\pm\gamma^5 \over 2}$, and
$W_\mu^{L,R}$ are general background {\it chiral} gauge fields.
Introducing some regularization the effective action can be defined;
$$
S_{eff} ( W^L, W^R ) = {\rm Tr} \ln
[ \gamma^\mu (\partial_\mu + i W_\mu^L(x) P_L + iW_\mu^R(x) P_R)  ] .
\eqno(2.2)$$
We denote the parity-odd part of $S_{eff}$ as $\Gamma$.
The gauge anomalies can be calculated through
$$ \delta \Gamma ( W^L, W^R) \equiv \Gamma ( W^L+\delta W^L,
W^R+\delta W^R) - \Gamma ( W^L,W^R) \eqno(2.3)$$
where
$$\cases {
\delta W^L(x)_\mu= -\partial_\mu\theta^L(x)+i[\theta^L(x), W^L_\mu(x)]
 \cr
\delta W^R(x)_\mu= -\partial_\mu\theta^R(x)+i[\theta^R(x), W^R_\mu(x)]
 \cr } \eqno(2.4) $$
and $\theta^{L,R}$ are infinitesimal gauge transformations of (1.2) so that
$O((\theta^{L,R})^2)$ terms  are neglected
in the definition of $\delta \Gamma$.

In order to calculate divergences of vector and axial-vector currents
we replace $W^{L,R}$ with
$$\cases {
\tilde W_\mu^L (x)  = W_\mu^L(x) + V_\mu (x) + A_\mu (x)  \cr
\tilde W_\mu^R (x)  = W_\mu^R(x) + V_\mu (x) - A_\mu (x)  \cr }
\eqno(2.5)
$$
where $V_\mu$ ( $A_\mu$ ) is an external U(1) vector( axial-vector) field.
They transform as
$$
\delta V_\mu = -\partial_\mu \theta^V (x)\qquad  {\rm and }\qquad
\delta A_\mu = -\partial_\mu \theta^A (x) \eqno(2.6)
$$
under infinitesimal gauge transformations $\theta^{V,A}$.
Since the U(1) vector and axial-vector {\it Noether} currents
are defined by
$$\eqalign{
J^\mu_V (x) & = {\partial S \over \partial V (x) } =
i\bar\psi\gamma^\mu\psi \cr
J^\mu_A (x) & = {\partial S \over \partial A (x) } =
i\bar\psi\gamma^\mu\gamma^5 \psi \cr
} \eqno(2.7) , $$
the expansion of $\delta \Gamma$ gives the following relation:
$$\eqalign{
\delta\Gamma (\tilde W^L, \tilde W^R) & = \delta\Gamma (W^L, W^R)
+\int dx \left[ \theta^V (x) \partial_\mu \vac{ J^\mu_V (x)} +
\theta^A (x) \partial_\mu \vac{ J^\mu_A (x)} \right] \cr
& +\int dx \left[ V_\mu (x)  \delta \vac{ J^\mu_V (x)}
+A_\mu (x) \delta \vac{ J^\mu_A (x)} \right] \cr
& \qquad + O( V^2, A^2, VA) \cr
} \eqno(2.8) $$
where
$$\eqalign{
\vac{J^\mu_V (x)} & = {\partial \Gamma\over \partial V_\mu (x) } (W^L, W^R) \cr
\vac{J^\mu_A (x)} & = {\partial \Gamma\over \partial A_\mu (x) } (W^L, W^R) \cr
\delta \vac{ J^\mu_V (x)} & = \delta
{\partial \Gamma\over \partial V_\mu (x) } (W^L, W^R) =
{\partial (\delta\Gamma) \over \partial V_\mu (x) } (W^L, W^R) \cr
\delta  \vac{ J^\mu_A (x) } & = \delta
{\partial \Gamma\over \partial A_\mu (x) } (W^L, W^R) =
{\partial (\delta\Gamma) \over \partial A_\mu (x) } (W^L, W^R) \cr
} \eqno(2.9) . $$
Here $\vac{\cal O} $ denotes the vacuum expectation value of $\cal O$ in
the presence of the background gauge fields $W^L$ and $W^R$.
{}From (2.8) we can easily obtain the divergences of U(1) vector and
axial-vector currents $ \partial_\mu \vac{J^\mu_{V,A} (x)}$ as well as
their variation $ \delta \vac{J^\mu_{V,A} (x)}$ under (1.2).
Finally it is noted that
$\Gamma$ term can be divided into 2 terms:
$$
\Gamma = \Gamma_{\rm pure} + \Gamma_{\rm local} \eqno(2.10)
$$
where $\Gamma_{\rm pure}$ has the non-local form of $W^L$ and $W^R$
while $\Gamma_{\rm local}$ only contains the local terms.
$\delta \Gamma_{\rm pure} (W^L,W^R)$ is the usual (consistent) gauge anomaly
and is regularization-independent.

\vskip 1cm
\centerline{III. \ Results of Dirac Type regularizations }
\vskip 0.5cm
The dimensional regularization[8] for the Dirac fermions
is defined by
$$
S_0+S_{reg} =\int d^D x \sum_{\mu=1}^D \bar\psi (x)
\gamma^\mu ( \partial_\mu + i W_\mu^L P_L + i W_\mu^R P_R ) \psi (x)
\eqno(3.1) $$
where $D=4-2\varepsilon$, the background chiral gauge fields $W_\mu^{L,R}$
are defined only in 4 dimensions and
we use the 't Hooft-Veltman definition[8] of $\gamma^5$ satisfying
$ \gamma^5 \gamma^\mu = - \gamma^\mu \gamma^5$ for $\mu = 1 \sim 4$ and
$ \gamma^5 \gamma^\nu = \gamma^\nu \gamma^5$  for $\nu = 5 \sim d$.
It is noted that because of this $\gamma^5$ property the left-handed part of
$\psi$ couples to the right-handed part of $\bar\psi$ through the term
$\sum_{\mu=5}^D\bar\psi \gamma^\mu \partial_\mu \psi$. Therefore we have to
assign which left-handed (Weyl) fermion couples to which right-handed (Weyl)
fermions in order to form one Dirac fermion and, for the one generation
standard model, we have to introduce a right-handed neutrino at the regularized
level. We call this type of regularization the Dirac type.

The lattice version of the Dirac type regularization is given by
$$\eqalign{
S_0 & = {1\over 2a}\sum_{x,\mu} \bar\psi (x) \gamma^\mu \left[
\{ U^L_\mu (x) P_L + U^R_\mu (x) P_R \}\psi (x+a\mu ) \right. \cr
& - \left.
\{ U^L_{-\mu}(x) P_L + U^R_{-\mu}(x) P_R \}\psi (x-a\mu ) \right] \cr
} \eqno(3.2a) $$
$$
S_{reg}  = -{r\over 4a}\sum_{x,\mu} \bar\psi (x) [ \{ U^L_\mu(x)  +
U^R_\mu(x)\}\psi (x+a\mu ) +\{ U^L_{-\mu}(x)+U^R_{-\mu}(x) \}\psi (x-a\mu )
-4\psi (x) ]
\eqno(3.2b) $$
where $a$ is the lattice spacing, $U_\mu^{L,R} = \exp [iaW^{L,R}_\mu ]$,
and $r$ in $S_{reg}$ is the Wilson-Yukawa coupling.
We need $U^L + U^R$ in $S_{reg}$ in order to use the formula for the
U(1) Noether currents in Sect. II.

For the dimensional regularization we obtain
$$\eqalign{
\delta \Gamma_{\rm pure}^{(dim)} (W^L, W^R) & = \int dx
\; i{\epsilon^{\mu\nu\alpha\beta}\over 24\pi^2}
{\rm tr} \left[\theta^L \partial_\mu (W_\nu^L\partial_\alpha W_\beta^L
+{i\over 2}W_\nu^LW_\alpha^LW_\beta^L ) \right. \cr
& \left. \phantom{= \int dx \; i{\epsilon^{\mu\nu\alpha\beta}\over 24\pi^2}}
-\theta^R \partial_\mu (W_\nu^R\partial_\alpha W_\beta^R
+{i\over 2}W_\nu^RW_\alpha^RW_\beta^R ) \right] (x)  \cr }
\eqno(3.3)$$
$$\eqalign{
\Gamma_{\rm local}^{(dim)} (W^L, W^R) & = \int dx\;
i{\epsilon^{\mu\nu\alpha\beta}\over 48\pi^2}
{\rm tr} \left[ (\partial_\mu W_\nu^L+\partial_\mu W_\nu^R
+{i\over 2}W_\mu^LW_\nu^L+{i\over 2}W_\mu^RW_\nu^R) \right. \cr
& \left. \qquad\times (W_\alpha^LW_\beta^R-W_\alpha^RW_\beta^L)
-{i\over 2}W_\mu^RW_\nu^LW_\alpha^RW_\beta^L \right] (x) , \cr
}  \eqno(3.4) $$
and for the lattice regularization
$$\eqalign{
\delta \Gamma_{\rm pure}^{(lat)} (W^L, W^R) & =
i\epsilon^{\mu\nu\alpha\beta} \int dx \;
{\rm tr} \left[\theta^L \partial_\mu \left\{
(I_1+I_2) W_\nu^L\partial_\alpha W_\beta^L
+i(I_1-I_2) W_\nu^LW_\alpha^LW_\beta^L \right\} \right. \cr
& \left.
-\theta^R \partial_\mu \left\{ (I_1+I_2)W_\nu^R\partial_\alpha W_\beta^R
+i (I_1-I_2) W_\nu^RW_\alpha^RW_\beta^R \right\} \right] (x)  \cr }
\eqno(3.5)$$
$$\eqalign{
\Gamma_{\rm local}^{(lat)} (W^L, W^R) & = i\epsilon^{\mu\nu\alpha\beta} \int dx
\; {\rm tr} \left[ \left\{ 2I_1/3 (\partial_\mu W_\nu^L+\partial_\mu W_\nu^R)
+i I_3 (W_\mu^LW_\nu^L+W_\mu^RW_\nu^R) \right\} \right. \cr
&\times \left. (W_\alpha^LW_\beta^R-W_\alpha^RW_\beta^L)
-i I_2 W_\mu^RW_\nu^LW_\alpha^RW_\beta^L \right] (x)  . \cr
}  \eqno(3.6) $$
Since
$$
I_1 = \int^\pi_{-\pi} d^4p {4r M S_\mu^2 - M^2 C_\mu\over F^3}C_\nu C_\alpha
C_\beta = {1\over 32\pi^2}  \eqno(3.7a)
$$
$$
I_2 = \int^\pi_{-\pi} d^4p {4r M^3 S_\mu^2 - M^4 C_\mu\over F^4}C_\nu C_\alpha
C_\beta = {1\over 96\pi^2} \eqno(3.7b)
$$
$$
I_3 = \int^\pi_{-\pi} d^4p {4r M S_\mu^2 - M^2 C_\mu\over 2F^4}S^2
C_\nu C_\alpha C_\beta = {1\over 96\pi^2} \eqno(3.7c)  ,
$$
where
$S_\mu = \sin p_\mu$, $C_\mu = \cos p_\mu$, $S^2 = \sum_\mu S_\mu^2$,
$M = r \sum_\mu ( 1- C_\mu )$ and $ F= S^2 + M^2 $,
we find that
$$ \delta \Gamma_{pure}^{(dim)} = \delta \Gamma_{pure}^{(lat)}
\qquad {\rm and} \qquad
\delta \Gamma_{local}^{(dim)} = \delta \Gamma_{local}^{(lat)}
\eqno(3.8)$$

{}From the expression above for $\Gamma$ and the formula in the previous
section
it is easy to see

\noindent (1) The U(1) axial-vector Noether current has an anomaly:
$$\eqalign{
\vac{ \partial_\mu J^\mu_A } & =  i{\epsilon^{\mu\nu\alpha\beta}\over 24\pi^2}
{\rm tr} \partial_\mu \left[ W_\nu^L\{ 2\partial_\alpha W_\beta^L
+i W_\alpha^LW_\beta^L + i W_\alpha^RW_\beta^R \} \right. \cr
& +\left. W_\nu^R\{ 2\partial_\alpha W_\beta^R
+i W_\alpha^LW_\beta^L + i W_\alpha^RW_\beta^R \}
+ 2 W_\nu^L\partial_\alpha W_\beta^R \right]  .
} \eqno(3.9) $$
For the QCD case ( $W^L = W^R = W$ ) this agrees with the well-known
result[9]:
$$
\vac{ \partial_\mu J^\mu_A } =  i{\epsilon^{\mu\nu\alpha\beta}\over 16\pi^2}
{\rm tr} G_{\mu\nu} G_{\alpha\beta}
\eqno(3.10) $$
where $G_{\mu\nu} = \partial_\mu W_\nu -\partial_\nu W_\mu + i [ W_\alpha ,
W_\beta ] $.
It is noted that the local term $\Gamma_{\rm local}$
is necessary to obtain the gauge invariant result (3.10)
with the correct normalization $1/(16\pi^2)$.

\noindent (2)
The axial-vector Noether current is {\it not} invariant under (1.2);
$$\eqalign{
\delta \vac{ J^\mu_A} & = i{\epsilon^{\mu\nu\alpha\beta}\over 24\pi^2}
{\rm tr} \left[ \partial_\nu (\theta^L-\theta^R) \{ \partial_\alpha (
W^L_\beta - W^R_\beta ) + i W^L_\alpha W^L_\beta -iW^R_\alpha W^R_\beta \}
\right.  \cr
& + (\theta^L-\theta^R) \left( i[W_\alpha^R,\partial_\nu W^L_\beta ] -
i[W_\alpha^L,\partial_\nu W^R_\beta ] \right.    \cr
& +\left.  \left. ( W^R_\nu W^R_\alpha W^L_\beta
+ W^R_\nu W^L_\alpha W^L_\beta - W^L_\nu W^R_\alpha W^R_\beta -W^L_\nu
 W^L_\alpha W^R_\beta \right)  \right]
} \eqno(3.11) .  $$
It is clear that the current is invariant under (1.2) only for
vector gauge theories such as QCD ( $\theta^L = \theta^R$ ).

\noindent (3)
Since the regularized action is manifestly invariant under the U(1) vector
transformation, the U(1) vector Noether current is conserved:
$$
\vac{\partial_\mu J^\mu_V} = 0  . \eqno(3.12)
$$
The fermion number $Q_F = \int d^3x J_V^0$ defined from the
vector Noether current is conserved.
This is the problem recently pointed out [2] for the lattice action.
However this vector Noether current also is {\it not} invariant
Under (1.2);
$$
\delta \vac{ J^\mu_V}  = i{\epsilon^{\mu\nu\alpha\beta}\over 8\pi^2}
{\rm tr}\partial_\nu [(\theta^L-\theta^R)(\partial_\alpha (
W^L_\beta + W^R_\beta ) + i [W^L_\alpha , W^R_\beta ] ) ]  .
\eqno(3.13) $$
Therefore, in the gauge non-invariant regularization
scheme, $Q_F$ does not correspond to the observed fermion number.
The following modified non-Noether currents:
$$\eqalign{
\tilde J^\mu_V & = J^\mu_V + K^\mu_V \cr
\tilde J^\mu_A & = J^\mu_A + K^\mu_A
}\eqno(3.14) $$
are gauge invariant under (1.2) [10]:
$$
\delta \vac{\tilde J^\mu_V } = \delta \vac{\tilde J^\mu_A} = 0  ,
\eqno(3.15) $$
where
$$\eqalign{
K^\mu_V  & = i{\epsilon^{\mu\nu\alpha\beta}\over 8\pi^2}
{\rm tr} \left[ W^L_\nu \{ \partial_\alpha
(W^L_\beta + W^R_\beta ) + {i2\over 3} [W^L_\alpha , W^L_\beta ] \} \right. \cr
& \left. -W^R_\nu \{ \partial_\alpha
(W^L_\beta + W^R_\beta ) + {i2\over 3} [W^R_\alpha , W^R_\beta ] \}
 \right] \cr
} \eqno(3.16) $$
$$
K^\mu_A  = i{\epsilon^{\mu\nu\alpha\beta}\over 24\pi^2}
{\rm tr} (W^L_\nu-W^R_\nu ) [\partial_\alpha (
W^L_\beta - W^R_\beta ) + i W^L_\alpha W^L_\beta -iW^R_\alpha W^R_\beta ]
. \eqno(3.17) $$
These modified currents give the desired anomalies:
$$
\partial_\mu \vac{\tilde J^\mu_V }
=  i{\epsilon^{\mu\nu\alpha\beta}\over 32\pi^2}
{\rm tr} \left[ G^L_{\mu\nu} G^L_{\alpha\beta} - G^R_{\mu\nu}
G^R_{\alpha\beta} \right]
\eqno(3.18a)$$
$$
\partial_\mu \vac{\tilde J^\mu_A}
=  i{\epsilon^{\mu\nu\alpha\beta}\over 32\pi^2}
{\rm tr} \left[ G^L_{\mu\nu} G^L_{\alpha\beta} + G^R_{\mu\nu}
G^R_{\alpha\beta} \right]
\eqno(3.18b) $$
where $G^L_{\mu\nu} = \partial_\mu W^L_\nu -\partial_\nu W^L_\mu +
i [ W^L_\alpha ,W^L_\beta ] $ and
$G^R_{\mu\nu} = \partial_\mu W^R_\nu -\partial_\nu W^R_\mu +
i [ W^R_\alpha ,W^R_\beta ] $.
The anomalies appear in the gauge-invariant non-Noether
currents, not in the Noether currents.
The gauge invariant fermion number is given by
$$
\tilde Q_F = Q_F + Q_B = \int d^3x\ \bar\psi\gamma^0\psi +
\int d^3x\ K^0_V  .
$$
Here $Q_B$ is equal to the winding number of the gauge field,
which is not conserved
in the presence of instantons.
In the Dirac type regularization scheme
the fermion number carried by the gauge field is changed by the
instantons while the fermion number carried by the fermion field
is always conserved.  Since the instantons induce no fermionic zero-modes
which violate fermion number in this regularization scheme,
no fermion-number violating vertex[1] is generated
from the fermion determinant.
If we can calculate the transition rate among the states with different $Q_B$
by some non-perturbative method, it directly gives the rate of fermion
number non-conservation without referring to fermionic matrix elements.

\vskip 0.5cm

Now we consider the case of the gauge invariant Dirac-type
regularization scheme. In this scheme, the relevant symmetry is
(1.5) and the conserved vector Noether current is invariant under (1.5).
Therefore, the fermion
number can not be violated in this formulation as pointed out in ref.[2].
Since the currents are always gauge invariant
due to the presence of scalar fields $g_L$ and $g_R$,
the gauge invariant fermion number is conserved in the
lattice chiral gauge theories with the Wilson Yukawa coupling.

\vskip 1cm
\centerline{IV. \ New Regularization for Weyl Fermions }
\vskip 0.5cm

The Dirac-type regularization used in the previous sections can not deal
with the one-generation standard model without right-handed neutrinos.
(It can deal with the one-generation standard model without right-handed
neutrinos. See ref.[11].)
In this section, we propose a new regularization scheme in order to deal with
a single left-handed (right-handed) fermion,
even at the regularized level.
We call this regularization scheme the Weyl type.  We combine the Weyl-type
regularization with the dimensional regularization and the
lattice regularization, though we can combine it with any other
regularizations such as the Pauli-Villars regularization.

The action for one left-handed fermion is
$$
S_0 + S_{reg}= \int d^D x\;
\sum_{\mu =1}^4 \bar\psi^L  \gamma_L^\mu ( \partial_\mu + ig W_\mu^L )
\psi^L +{1\over 2}\sum_{\mu =5}^d (\psi^L C^L \gamma_L^\mu\partial_\mu\psi^L
-\bar\psi^L\gamma_L^\mu \partial_\mu C^L\bar\psi^L )
\eqno(4.1)$$
for the dimensional regularization[11] and
$$\eqalign{
S_{reg}  & = -{r\over 4}\sum_{x,\mu}\left[ \psi^L(x)C^L\{
\psi^L(x+a\mu )+ \psi^L(x-a\mu ) - 2\psi^L(x)\} \right.  \cr
& - \left. \bar\psi^L(x)C^L\{ \bar\psi^L(x+a\mu )+
\bar\psi^L(x-a\mu ) - 2\bar\psi^L(x)\}  \right] \cr
} \eqno(4.2) $$
for the lattice regularization[12].
Here $\psi^L = P_L \psi$, $\gamma_L^\mu = \gamma^\mu P_L$, $C^L = C P_L$,
and $C$ is the charge conjugation
matrix which satisfies $ C \gamma^\mu C^{-1} = - (\gamma^\mu)^T $.
It is easy to see that the fermion number as well as the gauge symmetry
is violated by the Majorana type terms in $S_{reg}$ .

We obtain
$$
\delta \Gamma_{\rm pure} (W^L)  = \int dx
\; i{\epsilon^{\mu\nu\alpha\beta}\over 24\pi^2}
{\rm tr} \left[ \theta^L \partial_\mu \left( W_\nu^L\partial_\alpha W_\beta^L
+{i\over 2}W_\nu^LW_\alpha^LW_\beta^L \right) \right] \eqno(4.3) $$
and $\Gamma_{\rm local} (W^L) = 0 $ for both regularizations.
It is noted again that the parity-even terms are neglected here.
{}From the above expression,
we find that
the U(1) left-handed Noether current has non-zero divergence;
$$
\partial_\mu \vac{J^\mu_L} =  i{\epsilon^{\mu\nu\alpha\beta}\over 24\pi^2}
{\rm tr}\partial_\mu  ( W^L_\nu\partial_\alpha W^L_\beta + i
{1\over 2} W^L_\nu W^L_\alpha W^L_\beta )  ,
\eqno(4.4) $$
and the current is not gauge invariant;
$$
\delta \vac{ J^\mu_L} =  i{\epsilon^{\mu\nu\alpha\beta}\over 24\pi^2}
{\rm tr}\partial_\mu\theta^L  ( 2\partial_\alpha W^L_\beta + i
{1\over 2} W^L_\alpha W^L_\beta )  .
\eqno(4.5) $$
It is obvious that there is no mixing term between left- and right-
gauge fields and the right-handed Noether current becomes
$$
\partial_\mu \vac{J^\mu_R} =  -i{\epsilon^{\mu\nu\alpha\beta}\over 24\pi^2}
{\rm tr}\partial_\mu  ( W^R_\nu\partial_\alpha W^R_\beta + i
{1\over 2} W^R_\nu W^R_\alpha W^R_\beta )  .
\eqno(4.6) $$
It is noted that if we use this regularization for QCD,
the divergence of the U(1) axial-vector Noether current, $J^L_\mu - J^R_\mu$,
does not agree with the desired result, eq.(3.10) .
If we use this type of regularization for the one-generation standard model
without right-handed neutrinos, we obtain the anomaly for the baryon number
and the lepton number, as expected. However the result is not invariant
under (1.2) and, therefore, the left-handed fermion number,
$\int d^3 x J^0_L$, is not an
observable in the gauge non-invariant scheme.

The gauge invariant left-handed non-Noether current can be defined as
$$
\tilde J^\mu_L = J^\mu_L + K^\mu_L
\eqno(4.7) $$
where
$$
K^\mu_L  =  i{\epsilon^{\mu\nu\alpha\beta}\over 24\pi^2}
{\rm tr} W^L_\nu ( 2\partial_\alpha W^L_\beta + i
{3\over 2} W^L_\alpha W^L_\beta )  .
\eqno(4.8)$$
This modified current has the desired anomaly;
$$
 \partial_\mu\vac{\tilde J^\mu_L }
= i{\epsilon^{\mu\nu\alpha\beta}\over 32\pi^2}
{\rm tr } G^L_{\mu\nu} G^L_{\alpha\beta}.
\eqno(4.9)$$
Again the left-handed fermion number can be carried by the gauge field as
well as the fermion fields and both bosonic and fermionic left-handed fermion
numbers are violated by the instantons.
Finally we obtain
$$
\partial_\mu \vac{\tilde J^\mu_V}
=  \partial_\mu\vac{\tilde J^\mu_L+\tilde J^\mu_R } =
i{\epsilon^{\mu\nu\alpha\beta}\over 32\pi^2}
{\rm tr } \left[ G^L_{\mu\nu} G^L_{\alpha\beta} - G^R_{\mu\nu}
G^R_{\alpha\beta} \right]
\eqno(4.10a)$$
$$
\partial_\mu \vac{\tilde J^\mu_A}
=  \partial_\mu\vac{\tilde J^\mu_L-\tilde J^\mu_R } =
i{\epsilon^{\mu\nu\alpha\beta}\over 32\pi^2}
{\rm tr } \left[ G^L_{\mu\nu} G^L_{\alpha\beta} + G^R_{\mu\nu}
G^R_{\alpha\beta} \right]
\eqno(4.10b)$$
These anomalies in the Weyl type regularization are identical to
those in the Dirac type regularization, eqs.(3.18a-b).

{}From the result of sect. III. and IV., it is concluded that anomalies of the
gauge invariant non-Noether currents do not depend on types of
regularizations, Dirac type or Weyl type.
However so far we do not have regularization
schemes which have gauge-invariant vector and axial-vector {\it Noether }
currents.

\vskip 1cm
\centerline{V. \ Discussions}
\vskip 0.5cm

In this paper we have investigated the relation among regularizations,
anomalies and the fermion number non-conservation
in general chiral gauge theories.
The gauge invariance (1.2) play a crucial role in giving the
unique anomaly of the vector current, (3.18a) or (4.10a), which does not depend
on global symmetries of regularization schemes, and in inducing the
fermion number non-conservation.
Fields which carry the fermion number, however, depend on global symmetries
of regularization schemes. In particular we show that
bosonic fields can have non-zero fermion number non-perturbatively.
The structure of fermionic zero modes in the presence of instantons
also depends on global symmetries of regularization schemes.
In the Dirac type regularization scheme there is no fermionic
zero mode which violates fermion number.

The fermion number non-conservation can not occur in
lattice chiral gauge theories with the Wilson-Yukawa coupling
since the relevant gauge symmetry is different from (1.2).
Even if the fermion number is explicitly broken by
the (gauge invariant) Majorana Wilson-Yukawa coupling,
this explicit breaking  does not lead to the correct anomaly (4.10a).
(This is also true for the proposal of ref. [18].)
So far there is no satisfactory gauge invariant lattice formulation
which gives the correct anomaly of the vector current.

Finally we consider the lattice formulation by the Rome group[19] where a
dummy gauge singlet fermion $\chi$ is introduced to construct the Wilson term.
This formulation can deal with the left-handed (right-handed) fermion only
and is classified into the gauge non-invariant scheme.
The action for the regulator becomes,
$$\eqalign{
S_{reg} & = {1\over 2a}\sum_{x,\mu} \bar\chi^R (x) \gamma^\mu
\left[ \chi^R (x+a\mu )  - \chi^R (x-a\mu ) \right]
   -{r\over 2a}\sum_{x,\mu} \left[ \bar\chi^R (x)
\{ \psi^L (x+a\mu ) \right. \cr
& + \psi^L (x-a\mu )  -2\psi^L (x)\}
 + \left.  \bar\psi^L (x) \{ \chi^R (x+a\mu ) +
\chi^R (x-a\mu ) -2\chi^R (x) \} \right] \cr  }  \eqno(5.1) $$
for the left-handed fermion $\psi^L$.
The left-handed fermion number for the physical field $\psi^L$ is violated by
the Wilson term in $S_{reg}$.
This formulation is equivalent to the Dirac type regularization for a
Dirac field $ (\psi^L, \chi^R)$ with the gauge fields $( W^L_\mu, 0)$.
Thus, we obtain
$$
\delta \Gamma_{\rm pure} (W^L) = \int dx
\; i{\epsilon^{\mu\nu\alpha\beta}\over 24\pi^2}
{\rm tr} \left[\theta^L \partial_\mu (W_\nu^L\partial_\alpha W_\beta^L
+{i\over 2}W_\nu^LW_\alpha^LW_\beta^L ) \right] (x)
\eqno(5.2)$$
$$
\Gamma_{\rm local} (W^L, W^R)  = 0
 \eqno(5.3) $$
which is identical to the result of the Weyl type regularization,
(4.3).
Therefore we obtain the same results for the anomalies (4.4) and (4.5):
The U(1) left-handed Noether current for $\psi^L$ has non-zero
divergence and the current is non-invariant under (1.2). The invariant
left-handed non-Noether current has the desired anomaly (4.9)
and the fermion number is carried by both the fermion field and the
gauge field.

\vskip 1cm
\bigskip
\centerline{ ACKNOWLEDGMENTS}
\bigskip
We would like to thank
Drs. H. Kawai, K. Fujikawa,
J. Shigemitsu, R. Shrock and A. Ukawa for useful discussion and
Drs. H. Murayama and T. Yanagida for
useful comment on the formula in sect.II.

\vskip 1cm
\noindent {\bf References}
\vskip 0.5cm

\item{[1]}G. 't Hooft, \prl{37}{1976}{110};\pr{14}{1976}{3432}.
\item{[2]}T. Banks, \pl{272}{1991}{75}.
\item{[3]}P.D. Swift, \pl{145}{1984}{256}.
\item{[4]}J. Smit, {\it Acta Phys. Polon.} {\bf B17}(1986)531;
              \npp{4}{1988}{451}.
\item{[5]}S. Aoki, \npp{4}{1988}{479};\prl{60}{1988}{2109};
\item{   } \pr{38}{1988}{618};\npp{9}{1989}{584}.
\item{[6]}K. Funakubo and T. Kashiwa, \prl{60}{1988}{2133}.
\item{[7]}S. Aoki, \pl{247}{1990}{357}; \pr{42}{1990}{2806}.
\item{[8]}G. 't Hooft and M. Veltman, \np{44}{1972}{189}.
\item{[9]}S.L. Adler, {\it Phys. Rev.} {\bf 177}(1969)2426;
\item{   }J.S. Bell and R. Jackiw, {\it Nuovo Cimento} {\bf 60A}(1969)47.
\item{[10]}M.J. Dugan and A.V. Manohar, \pl{265}{1991}{137}.
\item{[11]}S. Aoki, ''Chiral Symmetry, fermion masses and global anomalies'',
UTHEP-224, talk given at {\it Hot Summer Daze Workshop},
BNL, New York, USA, 1991.
\item{[12]}S. Aoki, in Proceedings of 1991 International Symposium on
Lattice Field Theory, UTHEP-230, Nucl. Phys. B (Proc. Suppl.) in press.
\item{[13]}S. Aoki, I-H. Lee, and S.-S. Xue, {\it BNL Report} {\bf 42494}
(Feb. 1989); \pl{229}{1989}{79};
\item{   }I-H. Lee, \npp{17}{1990}{457}.
\item{[14]}S. Aoki, I-H. Lee, J. Shigemitsu, and R.E. Shrock,
\pl{243}{1990}{403}.
\item{[15]}S. Aoki, I-H. Lee, and R.E. Shrock,
\np{355}{1991}{383};
\item{   }S. Aoki, in {\it Strong Coupling Gauge Theories and Beyond},
eds. T. Muta and K. Yamawaki (World Scientific, 1991) p. 349;
\item{   }\npp{20}{1991}{589}.
\item{[16]}W. Bock, A.K. De, K. Jansen, J. Jersak, T. Neuhaus, and J. Smit,
\pl{232}{1989}{436}; \np{344}{1990}{207}; W. Bock and A.K. De,
\pl{245}{1990}{207}; A.K. De, \npp{20}{1991}{572};
W. Bock, A.K. De, C. Frick, K. Jansen, and T. Trappenberg, {\it HLRZ-91-21}.
\item{[17]}M. Golterman and D. Petcher,
\pl{247}{1990}{370}.
\item{[18]}E. Eichten and J. Preskill, \np{145}{1986}{179}.
\item{[19]}A. Borrelli, L.Maiani, G. Rossi, R. Sisto and M. Testa,
\pl{221}{1989}{360}; \np{333}{1990}{335};
\item{   }Y. Kikukawa, {\it Mod. Phys. Lett.}{\bf A7}(1992)871.

\vfill
\end